\documentstyle[art10,epsf]{article}
\begin{document}
 
\noindent
{\Large\bf
Particle Spectra from the ALCOR Model
}\\[3mm]
\hspace*{6.327mm}\begin{minipage}[t]{12.0213cm}{\large\lineskip .75em
J. Zim\'anyi\footnote{The talk was presented by J. Zim\'anyi at the
Strangeness'96 Workshop, 15-17 May 1996, Budapest, Hungary. It will be
published in the Proceedings of the Workshop,
Heavy Ion Physics {\bf 4} (1996) 15.}, T.S. Bir\'o, T. Cs\"org\H o, P. L\'evai
}\\[2.812mm]
Research Institute for Particle and
Nuclear Physics, \\
H-1525 Budapest, 114. POB 49, Hungary
\\[4.218mm]{\it
%
%
}\\[1.624mm]\noindent
{\bf Abstract.}
We introduce the Transchemistry Model as a dynamical extension
of the Algebraic Coalescence Rehadronization (ALCOR) model in
describing the hadronization of quark matter which is expected to
be produced in relativistic heavy ion collisions.
Results are presented for CERN SPS NA49 $Pb+Pb$ experiment
calculating hadron multiplicities and momentum spectra.
The freeze-out properties of different hadrons
are characterized by similar temperature,
density and flow profiles in cylindrically symmetric geometry.
\\[1.5mm]
\end{minipage}
\vspace{4 truemm}

\section{ Introduction}
 
The evolution of hadronic fireball was successfully described by
the hadrochemical model \cite{hchem} in medium energy heavy ion
collisions. This model is based on elementary processes in which
hadron-hadron collisions lead to the production of other type of
hadrons. The time-evolution of the hadronic system was described
by a set of coupled differential equations in which the
time-dependent gain and loss terms were determined by means of
microscopical cross-sections for the elementary processes.
The microscopical cross sections could be calculated in various
approximations (e.g. considering free cross section, or medium
effects or mean field, etc.), but the framework of the
hadrochemistry calculation remains unaffected.
 
On the basis of the success of the hadrochemistry the quarkochemistry
model \cite{qchem} was developed to describe the evolution of
quark matter produced at ultra-relativistic energies. Here the
constituents of the quark-gluon plasma became the ingredients of the
microscopical processes and the cross sections could be obtained in
various ways (e.g. from perturbative QCD).
 
In the present paper we introduce a model in which the elementary
processes contain the constituents of the {\bf semi-deconfined quark
matter} as incoming particles and the constituents of the hadron gas
as outgoing particles. In this way the
elementary processes of the phase transition from quark phase to hadron
phase can be modeled. This model, which is essentially an
extension of the ALCOR model \cite{ALCOR} \cite{ALCORS95}
 may be called {\bf transchemistry},
i.e. the chemical-like description of the phase transition.
 
\newpage

The basic assumption is the presence of the semi-deconfined quark matter.
This means that gluons already disappeared from the system (they decayed
or were absorbed), the main degrees of freedoms are quarks and antiquarks
in the deconfined phase, they
are dressed and ready for hadronization.
One can use various approximations in the
calculation of the cross sections for the elementary processes.
Even more, one can include new terms into the set of the coupled
differential equations to describe different collective
properties of the system, e.g. temperature, flow or the strength of
condensations, etc.
 
If we make a linearized approximation to the above mentioned set
of differential equations --- assuming  short time durations only ---
then we arrive to the Algebraic Coalescence Rehadronization
(ALCOR) model as presented in Ref.~\cite{ALCOR}.
The ALCOR model is very successful to describe the hadron yields
produced in different high energy heavy ion collisions \cite{ALCORS95}.
Both transchemistry and ALCOR model yield particle numbers in the
reaction volume $V$ which can be compared with the experimental
results. However a more complete comparison can be developed
if we know the full momentum distribution of the out-coming
hadrons. Since $\sigma$, the production (hadronization) cross section
has its maximum when the relative momentum of the participants is small
(see later eq.(\ref{S21}),
then we may assume that
{\bf the spectrum of the produced hadrons can be characterized
similarly as that of the quarks}.
Although a direct calculation of the
momentum spectra of the out-coming hadrons is in progress, which is
based on the microscopic coalescence applied in ALCOR
\cite{ALCORMOM}, in the present paper we use the approximation mentioned
before.
Since the hadrons will suffer inelastic
and elastic collisions during the time-evolution of the
hadronic fireball, one can assume that all hadrons can
be characterized with the same local temperature (or with the same
temperature profile in case of an extended particle system).
Because of the presence of a quark matter phase before
hadronization one can assume that the collective flow
profile will also be the same for the different hadron species.
Furthermore, we will assume, that all hadrons can be characterized
by the same density profile.
The collective behaviour combined with the particle
numbers obtained from transchemistry or ALCOR yield the necessary
constrains to obtain the momentum spectra for all hadrons if we
determined one of them (e.g. the spectrum of pions) from
the experimental results.
 
In Section 2 the basic equations of transchemistry will be summarized
and we show, how the former version of the ALCOR model can be
derived from the equations of transchemistry.
In Section 3 we display the ALCOR model,
emphasizing its main properties. In Section 4 the hadronization rate
will be introduced in the presence of flow which is a general result
for particle systems characterized by a flow pattern. In Section 5
we display the particle numbers obtained from ALCOR for the
Pb+Pb collision at SPS energy. We also calculate  the
fully equilibrated hadron gas scenario
and we discuss the main differences between these models.
In Section 6 we generate momentum distributions for pion
and proton spectra and compare with the experimental results
in the Pb+Pb collision.
 
\newpage
 
\section{ Transchemistry and ALCOR}
 
In this section
we will summarize the basic equations of transchemistry.
For simplicity let us
assume, that the net baryon charge of the system is zero and
only $\pi$ mesons are produced.
Thus the rate-equations for pion production from
semi-deconfined quark matter are the following:
\begin{equation}
{d \over {dt}}  \left[ V(t) n_\pi(t) \right] =
D^{(\pi)} \langle \sigma^\pi_{q{\overline q}} v \rangle_{(t)} ~\cdot ~ n_q (t)
\cdot n_{\overline q}
V(t) ~ - ~ n_\pi(t)\cdot V(t) \cdot \Gamma_\pi \\
\end{equation}
The rate equations can be expressed in terms of particle numbers, too.
\begin{equation}
{d \over {dt}} N_\pi (t) =
{ {D^{(\pi)} \cdot \langle \sigma^\pi_{q{\overline q}} v \rangle_{(t)} }
\over {V(t)} }
\cdot N_q (t) \cdot N_{\overline q} (t) - N_\pi(t) \cdot \Gamma_\pi ,
\end{equation}
One obtains similar equations for the quark and antiquark numbers:
\begin{eqnarray}
{d \over {dt}} N_q (t) &=&
 - ~ { {D^{(\pi)} \cdot \langle \sigma^\pi_{q{\overline q}} v \rangle_{(t)} }
\over {V(t)} }
\cdot N_q (t) \cdot N_{\overline q} (t) + N_\pi(t) \cdot \Gamma_\pi \\
{d \over {dt}} N_{\overline q} (t) &=&
 - ~ { {D^{(\pi)} \cdot \langle \sigma^\pi_{q{\overline q}} v \rangle_{(t)} }
\over {V(t)} }
\cdot N_q (t) \cdot N_{\overline q} (t) + N_\pi(t) \cdot \Gamma_\pi
\end{eqnarray}
Here $n_i(t)$ denotes the density of particle $i$, $V(t)$ is the
reaction volume  within which constant density was
assumed and $N_i(t)$ is the number of particle $i$.
The $\Gamma_\pi$ denotes the loss term for pions related to their
`decay' (or re-entering the semi-deconfined quark matter).
The $\langle \sigma^\pi_{q{\overline q}} v \rangle_{(t)}$ denotes
the momentum averaged elementary
hadronization rate for one hadronic degree of freedom
and $D^{(\pi)}$ is the spin degeneracy factor for the final hadron.
The temperature, density and flow profile for the quark distributions
inside the volume $V$ can be chosen in various ways. We will assume
that they are identical with the hadronic ones
obtained from the experiments.
The hadronization rate
can be calculated in many microscopic ways assuming different models:
quark coalescence \cite{ALCOR}, Nambu-Jona-Lasinio model \cite{KlevHuf},
or other effective quark models, e.g. \cite{Wetter}.
 
{}From eq.(2)-(3)-(4) one obtains that $N_\pi+ N_q=const.$ and
$N_\pi+ N_{\overline q}=~const.$, which is the number conservation for
the hadronization process $ q + {\overline q} \Longleftrightarrow \pi$.
If the microscopical hadronization mechanism is different,
e.g. $q+{\overline q} \Longleftrightarrow \pi + \pi$, then the above equations
will contain different coefficients. In its recent form 
transchemistry equations conserve
not only baryon numbers, but the numbers of the constituent quarks and
antiquarks.
 
In general case the particle  number production for meson $M$ can be
described as follows:
\begin{eqnarray}
{d \over {dt}} N_{M}(q_i,{\overline q}_j;t) &=&
{ {D^{(M)}\langle \sigma^{M}_{q_i{\overline q}_j} v \rangle_{(t)} }
\over {V(t)} }
N_{q_i} (t) N_{{\overline q}_j} (t)
- \sum N_{M}(q_i,{\overline q}_j;t)  \Gamma_{M}
\label{trans}
\end{eqnarray}
In the loss term the $\sum$ means all of the possible decay channels.
We plan to apply the transchemistry for fast hadronization processes,
in which case the deconfinement channels have minor role. Thus we will
neglect the decay channels in the followings.
 
If the hadronization starts at time $t_0$ and ends at time $t_0 + \tau$,
then one can obtain the produced hadron numbers from the
following integral:
\begin{eqnarray}
N_{M}(q_i,{\overline q}_j;t_0 + \tau) &=&
\int_{t_0}^{t_0 + \tau}
{ {D^{(M)} \langle \sigma^{M}_{q_i{\overline q}_j} v \rangle_{(t)} }
\over {V(t)} } \cdot
N_{q_i} (t) \cdot N_{{\overline q}_j} (t) \ dt \label{mesontrans}
\end{eqnarray}
 
In transchemistry various hadrons can be formed from the deconfined
quark matter. Meson formation is relatively simple, because
one step is enough to confine a quark-antiquark pair into meson-like hadron.
Baryon (and antibaryon) production might happen in three-body
interactions or sequentially via two-body interactions. In the latter
case diquarks (anti-diquarks) will be produced and the time-evolution
of their density should be followed also. The microscopical description
of baryon production is not a well-developed field, so we use
the simplest two-step processes to form baryons. In this case we need
to follow the production of diquarks and anti-diquarks also, as they are
carrying explicitly certain  degrees of freedom in transchemistry.

In case of  fast hadronization (e.g. an overcooled quark matter
is converted into hadrons \cite{CsorgCser}) the volume $V$ and the averaged
rate $\langle\sigma v\rangle$ will not change too quickly, thus one can
apply the
theorem of mean value of the integral calculus. For one integral this
transcription is exact.
The first term in eq.~(\ref{mesontrans}) ---
which contains the rate and the volume ---
remains approximately unchanged, but the time-dependent function of quark
numbers will be substituted by their mean value at time
$t^*_i$ ($t_0 < t^*_i < t_0+\tau$, where $\tau << t_0$):
\begin{eqnarray}
N_{M}(q_i,{\overline q}_j;t_0 + \tau) &=&
{ {D^{(M)} \langle \sigma^{M}_{q_i{\overline q}_j} v \rangle  \cdot \tau }
\over {V} }
\cdot N_{q_i} (t^*_i) \cdot N_{{\overline q}_j} (t^*_j)
\end{eqnarray}
 
One can approximate the particle numbers at $t^*$ by means of the
initial particle numbers: $N_{q_i}(t^*_i) \approx b_{q_i} N_{q_i} (t_0)$
and $N_{{\overline q}_j} (t^*_j) \approx
b_{{\overline q}_j}N_{{\overline q}_j} (t_0)$.
If the parameters $b_{q_i}$ and $b_{{\overline q}_j}$ were 
time-independent
then the above approximation can be useful to simplify the time-dependent
transchemistry equations:
\begin{eqnarray}
N_{M}(q_i,{\overline q}_j;t_0+\tau) &=&
{ {D^{(M)} \langle \sigma^{M}_{q_i{\overline q}_j} v \rangle  \cdot \tau }
\over {V} }
\cdot b_{q_i}N_{q_i}(t_0)  \cdot b_{{\overline q}_j} N_{{\overline q}_j} (t_0)
\label{mesonalc}
\end{eqnarray}
 
The complete set of time dependent differential equations of the type of
eq.~(\ref{trans}) preserve
all particle numbers which should be preserved --- as we mentioned earlier.
However, after the above linear
approximation the conserved particle numbers will not be preserved
automatically. Therefore we have to impose the conservation law of
particle numbers (namely quark and antiquark numbers)
while keeping the essence of the above set of equations.
In this way one arrives to the basic equations
of the phenomenological ALCOR model,
as a linearized, averaged special case of the transchemistry.
 
\section{ The ALCOR model}
 
The starting point of the ALCOR model \cite{ALCOR}\cite{ALCORS95}
is a semi-deconfined state in which only those quarks and antiquarks
are present in local thermal equilibrium
which form the final hadrons.
We also assume that only quarks and antiquarks are present, so their number
already accounts for those gluons which were fragmented earlier.
This assumption is  easily acceptable for  the heavy quarks  for
which any meeting  with their antiparticle  is improbable due  to
their small specific  density.  In the  case of light  constituent
quarks  (counted  after  gluon  fragmentation) the approximate
entropy conservation during  the rehadronization process  leads to
the conservation of  their numbers.  This  picture was  first
formulated in the framework  of the algebraic recombination  model
\cite{recom}.  In this model the  number of a given type of hadron
produced is
proportional to the product  of the numbers of  their constituting
quarks.   This   picture  was  developed   further  by
introducing gluon fragmentation \cite{gfrag}.
 We shall denote the number of produced $u,d,s$ quark pairs
--- just before the hadronization --- by $N_{u,pair}$, $N_{d,pair}$ and
$N_{s,pair}$, respectively.
The strangeness production factor, $g_S$ is defined as
$g_S = N_{s,pair} / ( N_{u,pair} + N_{d,pair})$.
 
The mesonic coalescence factors of the ALCOR model can be introduced
by means of eq.~(\ref{mesonalc}), namely
\begin{eqnarray}
C_{M}(q_i,{\overline q}_j) &=&
{ { \langle \sigma^{M}_{q_i{\overline q}_j} v \rangle  \cdot \tau }
\over {V} }
\end{eqnarray}
 
Thus the production of meson $M$ can be described as follows:
\begin{eqnarray}
N_{M}(q_i,{\overline q}_j) &=&
D^{(M)} C_{M}(q_i,{\overline q}_j)
\cdot b_{q_i}N_{q_i}  \cdot b_{{\overline q}_j} N_{{\overline q}_j}
\label{mesprod}
\end{eqnarray}
 
The baryonic (antibaryonic) coalescence factors can be defined and
calculated
as a simple sequential process, when baryons containing
three constituent quarks will be produced from
diquarks and quarks. The coalescence factors are the following:
\begin{eqnarray}
C_B(a,b,c)  &=&  g_B
{1\over 3}  \left\{ C_M(a,b)  C_M([a+b],c)  + \right.  \nonumber \\
 & \ & \ \ \left. +   C_M(a,c)  C_M([a+c],b) + C_M(b,c) C_M([b+c],a)
 \right\} .       \label{coalbar}
\end{eqnarray}
Here indices $a,b,c$ means the $q_i$ quarks for baryon production
or ${\overline q}_i$ antiquarks for antibaryon production.
The factor $g_B$ is the baryon suppression factor. Its physical meaning
can be connected to the sequential creation of the baryons.
We assume that the production of a baryon from three
quarks or an antibaryon from three antiquarks is a two-step
process leading through an intermediate diquark formation in the
corresponding color triplet or anti-triplet state which forms
the final baryon later
on together with a third quark or antiquark.
These diquarks must, however, be very unstable, short-lived
clusters, so they may decay before forming
the baryon.  We take into account this mechanism of baryon production
by introducing the phenomenological baryon suppression parameter,
 $g_B$.
Mathematically one can introduce the factor $g_B$ similarly as
the parameter $b_i$, namely approximating the time-dependent
integrals for baryon production with the substitution of mean
integral values. However, $g_B$ seems to be a well-behaving physical
parameter, so we will determine it from the experimental results,
similarly to $g_S$. We will assume that $g_B$ is flavour blind.
 
Now one can determine the numbers of the baryons and antibaryons:
\begin{eqnarray}
N_B(q_i,q_j,q_k)&=& D^{(B)} \, C_B(q_i,q_j,q_k) \cdot
b_{q_i}N_{q_i} \cdot b_{q_j} N_{q_j} \cdot
b_{q_k}N_{q_k} , \label{barprod} \\
N_{\overline B}({\overline q}_i,{\overline q}_j,{\overline q}_k)&=&
D^{({\overline B})} \,
C_{\overline B}({\overline q}_i,{\overline q}_j,{\overline q}_k) \cdot
b_{{\overline q}_i}N_{{\overline q}_i} \cdot b_{{\overline q}_j}
N_{{\overline q}_j} \cdot b_{{\overline q}_k}N_{{\overline q}_k} \ .
\label{abarprod}
\end{eqnarray}
 
In the above equations the spin degeneracies are also included,
$D^{(H)}=2S_H+1$. Here we shall consider
the simplest hadron multiplets conserving isospin symmetry. These are
the spin 0 and spin 1 meson octets together with the $\sigma$
and $\omega$ isospin singlet meson states. Furthermore the spin 1/2 octet
and spin 3/2 decouplet baryons and anti-baryons are also included.
 
Since we have introduced the numbers of different mesons,
baryons and antibaryons
produced from the quark matter, one has to put up the final
set of equations, which will ensure the conservation of
 the numbers of different flavours
(for each quark and each antiquark flavor):
\noindent
\begin{eqnarray}
& & N_{q_i}=
  \sum_H \sum_{j=1}^{N_f}
D^{(M)} C_{M}(q_i,{\overline q}_j)
\cdot b_{q_i}N_{q_i} \cdot 
b_{{\overline q}_j} N_{{\overline q}_j} \nonumber \\
& & +  \sum_H \sum_{j=1}^{N_f}  \sum_{k=j}^{N_f} ( 1 + \delta_{i,j} +
\delta_{i,k} )
D^{(B)} \, C_B(q_i,q_j,q_k) \cdot
b_{q_i}N_{q_i} \cdot b_{q_j} N_{q_j} \cdot b_{q_k}N_{q_k} \hspace{0.5cm}
\label{S4} \\
& &N_{{\overline q}_i}=
 \sum_H \sum_{j=1}^{N_f}
D^{(M)} C_{M}({\overline q}_i,q_j)
\cdot b_{{\overline q}_i}N_{{\overline q}_i} 
\cdot b_{q_j} N_{q_j} \nonumber \\
& & + \sum_H \sum_{j=1}^{N_f}  \sum_{k=j}^{N_f} (1+\delta_{i,j}+\delta_{i,k})
D^{(\overline B)} C_{\overline B}
({\overline q}_i,{\overline q}_j,{\overline q}_k) \cdot
b_{{\overline q}_i}N_{{\overline q}_i} \cdot
b_{{\overline q}_j}N_{{\overline q}_j} \cdot
b_{{\overline q}_k}N_{{\overline q}_k} . \label{S5}
\end{eqnarray}
 
This set of equations determines uniquely the normalization factors, $b_i$.
The number of independent equations in eqs.(\ref{S4}),(\ref{S5}) is
equal to the number of independent $b_{q_i}$ and
$b_{{\overline q}_i}$ factors. Note, that these $2N_f$ quantities are not
adjustable parameters of the model, but they are determined
by the flavour conservation.
 
Furthermore, introducing the factors $b_{q_i}$ and $b_{{\overline q}_i}$
we have constructed a scale invariant description for the
particle production. Namely, the physical quantities can be transformed
in the following way:
\begin{eqnarray}
C_{M}({\overline q}_i,q_j)
~ &\longrightarrow& ~ \lambda_i~{\overline \lambda}_j~
C_{M}({\overline q}_i,q_j),
    \nonumber\\
C_B(q_i,q_j,q_k) ~ &\longrightarrow& ~\lambda_i ~\lambda_j ~\lambda_k ~
C_B(q_i,q_j,q_k), \nonumber \\
C_{\overline B}({\overline q}_i,{\overline q}_j,{\overline q}_k)
~ &\longrightarrow& ~
    {\overline \lambda}_i~{\overline \lambda}_j~
    {\overline \lambda}_k ~
C_{\overline B}({\overline q}_i,{\overline q}_j,{\overline q}_k),
\nonumber\\
  b_{q_i}~ &\longrightarrow&~  b_{q_i} / \lambda_i ,\nonumber\\
  b_{{\overline q}_i}~ &\longrightarrow& 
  b_{{\overline q}_i} / {\overline \lambda}_i
\end{eqnarray}
 
Substituting the results of this transformation into eqs.
(\ref{mesprod}), (\ref{barprod}), (\ref{abarprod}), one can see that
the numbers of produced hadrons,
$N_{M}(q_i,{\overline q}_j)$, $N_B(q_i,q_j,q_k)$ and
$N_{\overline B}({\overline q}_i,{\overline q}_j,{\overline q}_k)$
remain unchanged. Thus even if we miss a constant factor in
the microscopical hadronization coefficients, then
one can obtain the particle numbers
correctly.

\section{ Hadronization Rate in the Presence of Flow}
 
In the presence  of a
flow we consider the relativistic J\"uttner distribution
\begin{equation}
f(x,p)  = e^{- \beta \, p \cdot u(x) }      \label{S12}
\end{equation}
 
  \noindent with $u_{\mu}(x)$ being  the local four-velocity  of
the flow.   Let us restrict ourselves in the following
to a scaling longitudinal flow \cite{bjorken}.
In this case the thermal averaging
of the rate can be interpreted only locally: the earlier rate equations
has to be reinterpreted in terms of reacting
components from different coordinate rapidity ranges
$\eta_a$ and $\eta_b$ respectively
 
\begin{equation}
{1 \over V}  {d \over {d\tau}} \left( V n_h \right) = \int
d\eta_a \, d\eta_b \, \langle \sigma v \rangle_{ab} n_a n_b. \label{S14}
\end{equation}
 
\noindent
Clearly the total volume of the fireball increases due to a
longitudinal scaling expansion like
$ V  = \tau \pi R^2 \,  \Delta \eta$
with total coordinate rapidity extension $\Delta \eta$ and
transverse radius $R$. Assuming a space-time rapidity plateau
for a finite interval in the number
distribution of the reacting and produced components
we arrive at a modified rate equation
 
\begin{equation}
\tau   {d \over {d\tau} } N_h  =  {\lambda^* \over {\pi R^2} } N_a
N_b \ .  \label{S16}
\end{equation}
 
\noindent Here
 
\begin{equation}
 \lambda^*  = \int d(\eta_a - \eta_b) \, \langle \sigma v \rangle \, \Theta
\bigl( (p_a-p_b) \cdot  (x_a - x_b) \bigr)     \label{S17}
\end{equation}
 
\noindent is  the total  rate of reactions between all possible coordinate
rapidity cell pairs.  The constraint $(p_a - p_b) \cdot  (x_a - x_b) \ge
0$ reduces to the requirement that the relative velocity vector is
oriented opposite to the relative position vector in the center
of mass system of the colliding components 'a' and 'b'.
This constraint is necessary because  only those particles
collide which have a relative velocity  pointing towards each other.
 
The inclusion of this constraint after some algebraic manipulation
leads to the general expression of the averaged rate
\begin{equation}
 \lambda^*  = {{\beta \int d \sqrt{s} \,\, \sigma(s)
\lambda^*_{ab}(s)
G(\beta \sqrt{s} ) } \over { 8m_a^2m_b^2K_2(\beta m_a) K_2(\beta
m_b) } }    \label{S18}
\end{equation}
 
\noindent where
$ \lambda^*_{ab}(s)  = \left[ s-(m_a+m_b)^2 \right] \left[
s-(m_a-m_b)^2 \right]$
 and $G(\beta \sqrt{s})$ is the thermal weight
factor.  In the presence of longitudinal
scaling  flow  the  latter  can  be  written  as  an integral over
relative coordinate rapidity $\eta = (\eta_a - \eta_b)$
 
\begin{equation}
 G(\beta \sqrt{s})  = \int_{-\infty}^{\infty} \, d\eta \,
\sqrt{s} \, \,
{ {
K_0\left(\beta \sqrt{s}\, {\rm ch} |\eta| \right) -
K_0 \left( \beta \sqrt{s}\, {\rm ch} |\eta|  \, + \,
\beta {{\sqrt{\lambda^*_{ab}(s)}}\over {2 \sqrt{s}}} {\rm sh}|\eta| \right)}
\over {\beta \sqrt{\lambda^*_{ab}(s)}\, {\rm ch} |\eta|\, {\rm sh} |\eta|}}.
\label{S19}
\end{equation}
Investigating the integrand of eq.(\ref{S19}) one observes a
finite width,
$\delta \eta$, in the relative coordinate rapidity $\eta$.
(This quantity, $\delta \eta$, is
equal to the relative flow rapidity because of the Bjorken scaling
assumption.) The finite width depends
on temperature $T$, on the particle
rest masses $m_a, m_b$, and on the considered energy scale
$\sqrt{s}$ in a complicated manner.
 
The most  uncertain ingredient  of the  ALCOR
had\-ro\-ni\-za\-tion model is the had\-ro\-ni\-za\-tion cross section
in eq.~(\ref{S18}).  We will consider
an analogy with the $p + A \rightarrow  d
+  (A-1)$  nuclear  rearrangement  (pick-up) reaction leading to
deuteron formation \cite{schiff}.  Assuming  a Coulomb-like potential
in the final ($q{\overline q}$) state
a fusing cross section can  be derived  from this  analogy:
\begin{equation}
 \sigma  = 16 m_h^2 \sqrt{\pi} \rho^3 { {\alpha^2 a } \over
{\left( 1 + (k \,  a)^2 \right)^2 }  } \label{S21}
\end{equation}
Here $m_h$ is the rest mass of the meson, while  $a
= 1  / (m_{ab}  \alpha) $  is the  Bohr radius  of the  bound $q
\overline{q}$ state in a $V(r) = - \alpha / r$ Coulomb potential
with $m_{ab}$ being the reduced  mass of particles 'a' and  'b'.
In our calculation we took $\alpha = 0.46$ and $T = 200$ MeV temperature
in estimating hadronization cross sections at CERN SPS  energy.
The factor $\rho  = 0.3$ fm  occurring in eq.(\ref{S21})
accounts for the medium influencing the hadron  formation and
was  taken  to  be  equal  to  the  Debye  screening  length  in
quark-gluon  plasma  at  the  above  temperature.   Finally  $k$
occurring in eq.(\ref{S21})  is the magnitude  of the relative  momentum
vector of particles 'a' and 'b' measured in their center  of
mass system $k = \sqrt{\lambda^*_{ab}(s)} / 2\sqrt{s}$.
 
%
 
\section{ Results for Pb+Pb collision from ALCOR}
 
Well armed with the above theoretical considerations the
redistribution of different flavor quarks and antiquarks into all
possible hadrons can be calculated in the ALCOR model using
practically only three parameters: i) the total number of
quark-antiquark pairs, $N_{tot,pair} =
N_{u,pair}+ N_{d,pair} + N_{s,pair} $,
which can be determined from the measured
total charged multiplicity, ii) the parameter $g_B$ controlling
the baryon formation and iii) the strangeness
production factor $g_S$.
The dependence of the final results on the other parameters
(e.g. $T$, $\alpha$) within their physically acceptable interval is small.
 All further
results, such as the number of hyperons, kaons, etc. are predictions
of the ALCOR model.
 
In Ref.~\cite{ALCORS95} we presented our results for the
S+S collisions at 200  GeV/nucleon bombarding energy.
We took into account that
the number of participant nucleons was measured to be
$N_{partic}^{SS} = 51$ \cite{particip}.  We used the parameters
$N_{tot,pair}^{SS}=158.1$ with $g_S=0.255$ for the strangeness production,
and $g_B=0.04$ for the baryon formation. The obtained particle
numbers fitted the experimental results excellently.
 
{}From this experience we conjectured 
that in S+S collision the rehadronization
process is {\bf locally quick}: the quarks and antiquarks (including the
fragmented   gluons)  appear   in   hadrons   according  to simple
kinematic rules and production branching ratios.
This finding, however, does not settle the question whether
quark  matter   with  any  degree   of  collectivity has been formed
or other hadronization  processes  including  strings
or color ropes \cite{rope} are  the source of this amount of quarks.
Local thermal distribution  of the  different  quark flavors  and the
presence of a  longitudinal flow cannot, on the other hand, be excluded
on the basis of this experimental data.
 
To obtain results for A+A collisions we assume a scaling connected
to the participant nucleon number for the produced
quark-antiquark pairs:
\begin{equation}
N_{tot,pair}^{AA} (\sqrt{s}) =
\left( { N_{partic}^{AA} \over N_{partic}^{SS}}\right)^\alpha
N_{tot,pair}^{SS} (\sqrt{s}), \label{S22}
\end{equation}
\noindent where $N_{partic}^{AA}$ are the number of participant nucleons in the
A+A collision.
The scaling exponent $\alpha$ may have the value $\alpha=1$,
or, for more collective production processes one expects $\alpha > 1$.
Furthermore we shall use the  values for $g_B$ and $g_S$ obtained above.
 
At the extrapolation of
the  total  number  of  produced quark-antiquark pairs from the S+S collision
one has to consider an energy scaling. We will assume a logarithmic one:
\begin{equation}
{{N_{tot,pair}^{AA} (\sqrt{s_1}) } \over
 {N_{tot,pair}^{AA} (\sqrt{s_2}) }}  =
{ {\ln{\sqrt{s_1}}} \over {\ln{\sqrt{s_2}}} }
\end{equation}
In our case, considering 160 GeV/nucleon collision energy for the
Pb+Pb collision, this rescaling yields
an $\approx 4 \%$ correction
for the $N_{tot,pair}^{PbPb}$.

\begin{table}[tbp]
\begin{center}
\begin{minipage}[tbp]{11.054cm}
{\small {\bf Table 1.}
{Particle numbers:
prediction of ALCOR model for the Pb+Pb collision
and the results of full equilibrium
thermal models, `Thermal I-II'. \\ } }
\end{minipage}
\begin{tabular}{||c||c|c|c||}    \hline
\hline {\bf Pb+Pb} &
 {\bf ALCOR} & {\bf Thermal I.} & {\bf Thermal II.} \\
\hline
\hline
 $h^{-}$&  730.41  & 798.72& 839.44 \\
\hline
\hline
 $\pi^+$&  603.87 & 603.97 & 603.54 \\
\hline
 $\pi^0$&  618.95 & 620.70 & 619.16 \\
\hline
 $\pi^-$&  634.68 & 637.91 & 635.31 \\
\hline
 $K^+$  &  \ 84.15 & 174.57 & 205.68 \\
\hline
 $K^0$  &  \ 84.15 & 177.74 & 211.29\\
\hline
 ${\overline K}^0$& \ 41.65  &103.10 & 119.46 \\
\hline
 $K^-$  &  \ 41.65  & 104.86 & 122.69 \\
\hline
 $K^0_{S}$&  \ 62.90  &  140.42 & 165.38 \\
\hline
\hline
 $p^+$  &  170.90  & 141.58& 129.97 \\
\hline
 $n^0$  &  188.57  & 148.20& 133.44 \\
\hline
 $\Sigma^+$&  \ 12.86  & \ 24.21& \ 41.27 \\
\hline
 $\Sigma^0$&  \ 13.63  & \ 25.04& \ 42.37 \\
\hline
 $\Sigma^-$&  \ 14.43  & \ 25.89& \ 43.51 \\
\hline
 $\Lambda^0$& \ 68.20  & \ 55.87& \ 42.37 \\
\hline
 $\Xi^0$&  \ \ 8.82  & \ 10.92 & \ 11.99 \\
\hline
 $\Xi^-$&  \ \ 8.89 & \ 11.09  & \ 12.31 \\
\hline
 $\Omega^{-}$&  \ \ 1.48  &  \ \ 1.56 & \ \ 2.92 \\
\hline
\hline
 ${\overline p}^-$&  \ 25.07 & \ 13.28 &\ 14.57 \\
\hline
 ${\overline n}^0$&  \ 25.07 & \ 12.68 &\ 14.19 \\
\hline
 ${\overline \Sigma}^-$&  \ \ 4.18  & \ \ 4.10 &\ \ 8.10 \\
\hline
 ${\overline \Sigma}^0$&  \ \ 4.18  & \ \ 3.96 &\ \ 7.89 \\
\hline
 ${\overline \Sigma}^+$&  \ \ 4.18  & \ \ 3.83 &\ \ 7.68 \\
\hline
 ${\overline \Lambda}^0$&  \ 20.93  & \ \ 8.84 &\ \ 7.89 \\
\hline
 ${\overline \Xi}^0$&  \ \ 5.98  & \ \ 3.10 &\ \ 3.91 \\
\hline
 ${\overline \Xi}^+$&  \ \ 5.98  & \ \ 3.05 &\ \ 3.81\\
\hline
 ${\overline \Omega}^{+}$& \ \ 2.20&\ \ 0.77 &\ \ 1.59 \\
\hline
\hline
\end{tabular}
\end{center}
\end{table}

Now we can make prediction for  the
total particle production in the Pb+Pb collision.
The number of participant nucleons is
$N_{part}^{PbPb} = 390 \pm 10$
obtained from Monte-Carlo simulations.
Using the mean value and a linear scaling ($\alpha=1$)
 one obtains $N_{tot,pair}^{PbPb}=1164$. We keep the strangeness
production and baryon formation factors fitted for S+S collision,
$g_S=0.255$, $g_B=0.04$.
Table 1 shows
our prediction for the particle numbers
together with the results of two thermal equilibrium models,
Thermal I and Thermal II.
In Table 2 some characteristic particle ratios are displayed for the
three models.
 
\newpage

In the thermal models we assume a homogeneous system in full chemical
equilibrium at different level. In model 'Thermal-I' the same
hadron multiplets are assumed to be in equilibrium, which are
populated in the ALCOR model, and then the short lived resonances
are allowed to decay into the stable particles shown in the Tables.
On the other hand, for the model 'Thermal-II' we assumed, that the
final (stable) particles are in chemical equilibrium.
In both cases the values were determined to get
approximately the same $\pi^+$ number as in ALCOR at a given
temperature, $T=165 \ MeV$. The chemical potential for strange
quarks were determined to reproduce strangeness neutrality.
The light quark chemical potential was obtained from the total baryon number
of the system. 
 
\begin{table}[tbp]
\begin{center}
\begin{minipage}[tbp]{11.054cm}
{\small {\bf Table 2.}
{Particle ratios: prediction of ALCOR model for the Pb+Pb collision
and the results of full equilibrium
thermal models, `Thermal I-II'. \\ } }
\end{minipage}
\begin{tabular}{||c||c|c|c||}    \hline
\hline {\bf Pb+Pb} &
 {\bf ALCOR} & {\bf Thermal I.} & {\bf Thermal II.} \\
\hline
\hline
 $p^+/\pi^+$& 0.28 & 0.23 & 0.21\\
\hline
 ${\overline p}^-/p^+$&  0.14& 0.09 & 0.11  \\
\hline
 ${\overline p}^-/\pi^-$& 0.04 & 0.02 & 0.02  \\
\hline
 $K^+/\pi^+$&  0.14& 0.29  & 0.34 \\
\hline
 $K^+/K-$&  2.02& 1.66 & 1.67  \\
\hline
 $K^0_S/\Lambda^0$&  0.92& 2.51 & 3.90 \\
\hline
 ${\overline \Lambda}^0/{\overline p}^-$&  0.83
& 0.66 & 0.54  \\
\hline
 ${\overline \Lambda}^0/\Lambda^0$&  0.30
& 0.16 & 0.18  \\
\hline
 $\Xi^{-}/\Lambda^0$& 0.13& 0.19& 0.29 \\
\hline
 ${\overline \Xi}^{+}/{\overline \Lambda}^0$
  & 0.28& 0.34& 0.48 \\
\hline
 $\Xi^{-}/{\overline \Xi^{+}}$ & 1.48 &  3.62 & 3.22  \\
\hline
 ${\overline p}^{-}/K^-$&$0.60$ &
   $0.13$& $0.12$ \\
\hline
 $\Lambda^{0}/K^-$&$1.64$ &
   $0.53$& $0.35$ \\
\hline
\hline
\end{tabular}
\end{center}
\vskip -1cm
\end{table}

   Comparing the results from  ALCOR with the results from the
thermal model, we find  that, surprisingly enough, they are not
far from each other. The greatest deviation is about a factor
of five in some of the particle ratios (see e.g. 
${\overline p}^-/K^-$).
 
 This similarity is due, probably, to the fact that, i) we have a
large number of resonances; ii) the spin degeneracy factor appears
in both models; iii) the conservation rules lead to strong
restrictions.
 
 Besides the similarities there are systematic differences between
the two models.  The strange baryon to strange meson ratio in the
thermal model is systematically smaller than in the ALCOR results.
This can be interpreted as the effect of mass differences, which
leads to an exp(-m/T) suppression in the thermal case, while it is
not present in ALCOR.  Thus this ratio may indicate that how
strong was the hadrochemical evolution in the final hadronic
phase.

\newpage
 
\section{ Momentum distribution}

In the ALCOR  model we have  a microscopical dynamical  prescription for
the process  of quark  coalescence into  hadrons.  Furthermore  we assumed a
given  momentum  distribution  for  the  constituent  quarks.   Thus the
momentum  distribution  of  the  produced  hadrons  can  be  calculated.
However, this  is a  somewhat tedious  procedure.  On  the other  hand we
observe, that the capture cross section ( eq.(\ref{S21}) ) has its maximum at
zero  relative  quark  momentum.   Therefore  one  may  assume, that the
hadrons, just after their  production will have a  momentum distribution
similar to that of  the quarks ( eq.(17)  ).  In this chapter,  however,
for the hadrons  we shall use  Fermi and Bose  distributions, and  a
transverse flow is added, too  (see eq.(28) below).  We assume,  that, due
to the elastic collisions, all the hadrons will have the  same density,
flow and temperature profiles.   However, the particle composition  will
remain the same as it was at the hadronization.  Thus we have chemically
not equilibrated system,  where all hadrons  will have its  own chemical
potential, corresponding to  the partial density  of this species.   The
value of the  temperature, $T$, the transverse radius, $R$, the radial flow
velocity at the surface and the proper time, $\tau$, are determined by
fitting the measured rapidity and transverse mass distribution.  However,
due to our assumptions, the same set of parameters must be used for  all
type of particles.  Thus the  existence of such a self-consistent  fit is
not trivial!  From the
momentum distributions and  the volume we  can calculate the  entropy of
the system, too.
 
In general the invariant momentum distribution will be the
following:
\begin{equation}
 E {  {d^3N}\over {d^3p} } =
 {1 \over {(2 \pi \hbar c)^3 } } \int {g_{deg} \over
{ e^{( p^\nu u_\nu - \mu )/T}+ K }}  p^\nu  d \sigma_\nu
\end{equation}
We will keep the quantum statistics, so for mesons we have $K=-1$, for
baryons and antibaryons $K=1$. Later we will see the importance of
quantum statistics for pions. For correct freeze-out calculation we
need to introduce the Cooper - Frye pre-factor
$p^\nu d\sigma_\nu$ \cite{CoopFrye}.
Let us consider a cylindrically symmetric geometry and assume
freeze-out at constant proper time $\tau$. For such a freeze-out
hypersurface one obtains
\begin{equation}
 p^\nu d\sigma_\nu =   m_t \  r dr \ \tau cosh( y -
\eta) \ d \eta \ d\phi_r .
\end{equation}
 
Furthermore, let us assume a longitudinal Bjorken flow and
a linear self-similar transverse flow,
\begin{equation}
 u_\nu= [cosh( \eta ) cosh( \Theta ), sinh(\eta) cosh(\Theta ),
        sinh(\Theta) cos(\phi_r),  sinh(\Theta) sin(\phi_r) ]
\end{equation}
where the transverse flow is defined as
$v_t=\tanh(\Theta)$. The transverse radius depends on the
proper time during expansion, namely it is $R(\tau)$.
 
 \newpage

After trivial calculation one obtains the invariant momentum spectra:
\begin{equation}
 { {dN}\over {m_t dm_t dy} } =
 {{\tau ~ g_{deg}} \over {2 \pi (\hbar c)^3 } } \int_0^{R(\tau)} r dr
 \int_{\eta_{min}}^{\eta_{max}} d\eta
 \int_0^\pi d\phi  { {m_t \cosh (y-\eta}) \over
{ e^{( p^\nu u_\nu - \mu )/T}+ K }}
\label{freeze}
\end{equation}
where $\phi = \phi_r -\phi_p$ and
$p^\nu u_\nu  = m_t \cosh(\Theta) \cosh(y-\eta) - p_t
\sinh(\Theta)\cos\phi $.
 
The calculation of particle spectra is based on the eq.~(\ref{freeze}),
where the thermodynamical parameters depend on the radius,
$T(r)$ and $\mu_(r)$. Considering Bjorken scaling, we neglect here
longitudinal dependence, only the finite longitudinal size
of the fireball is included, $[z_{min}, z_{max}]=
[\tau \cdot \sinh \eta_{min}, \tau \cdot \cosh \eta_{min}]$.

If we investigate the radial dependence of the particle numbers,
we need to introduce the following distribution function.
\begin{equation}
 { {dN}\over {r dr} } =
 {{\tau ~ g_{deg}} \over {2 \pi (\hbar c)^3 } }
 \int_{\eta_{min}}^{\eta_{max}} d\eta
 \int_{-\infty}^{+\infty} dy
 \int_m^{+\infty} m_tdm_t
 \int_0^\pi d\phi  { {m_t \cosh (y-\eta}) \over
{ e^{( p^\nu u_\nu - \mu )/T}+ K }}
\label{freezer}
\end{equation}
We assume that these radial distributions are similar for different
particle species at freeze-out time $\tau_f$.
 
The total energy of the system can be obtained from eq.(30) by
multiplying the integral with $E=m_t \cosh y$ and integrating
over $r$.
 
Before comparing the experimental results and the
calculated particle spectra, let us summarize shortly the
possible values of different parameters.
For  the temperature profile
 no longitudinal dependence was considered,
the radial dependence was chosen to be quadratic 
or higher, $T(\eta,r,\tau) = T_0(\tau) \cdot [1-(r / R(\tau))^\gamma$],
$T_0(\tau_f) = 150-180 \ MeV$, $\gamma \geq 2$. There is a
longitudinal Bjorken scaling and for the transverse flow we will use
the parameterization
$v_T(r,\tau)=\tanh (\Theta(r,\tau))$, where
 $\sinh(\Theta(r,\tau)) = w_T(\tau) \cdot r/R(\tau)$ and on the
 radial surface at freeze out time it is
 $v_T(R(\tau_f),\tau_f) = w_T(\tau_f) / \sqrt{w_T^2(\tau_f) + 1 } \ < \ 1$.
 The transverse radius, $R(\tau_f)$,
 the minimal and maximal value of the space-time rapidity, ($\eta_{min}, \
\eta_{max}$) and the proper time of the freeze-out $\tau_f$ are free
parameters ($z_{min} = \tau \sinh \eta_{min}$,
$z_{max} = \tau \sinh \eta_{max}$).
Chemical potentials, $\mu(r)$  will be calculated from dN/rdr.
 
Fig.~1 and 2 display the calculated $\pi^-$ and proton spectra  at
$\tau_f = 6 \ fm$,
$T_0(\tau_f) = 165 \ MeV$, $\gamma=2$,
$\eta_{min,max}=\pm 1.75$ (which means
longitudinal size $|z|=\pm 16.75 \ fm$),
$R_T(\tau_f)= 14 \ fm$, $v_T(R(\tau_f))=0.77$. 
In Fig.~2 the full lines indicate Bose and Fermi distributions, the  dotted
line indicates results for pion with Boltzmann distribution.
The transverse momentum distribution of pions shown in Fig. 2.
suggests that the Bose statistics plays a role in the enhancement
of the low 
$m_t$ part of the pion spectra. Such an effect was suggested
earlier for the heavy ion collisions at Bevatron  \cite{PRL} . If we
increase the freeze out volume by a factor
of two, the pion density decreases, and the low
$m_t$  enhancement
disappears.
 
The total entropy and energy of the hadron system were discussed
in Ref.~\cite{Par96}, where it was obtained 
$S/N_B = 30$ for the specific entropy in the above parametrized
hadron gas.
 
\newpage
\begin{center}
\vspace*{14cm}
\includegraphics{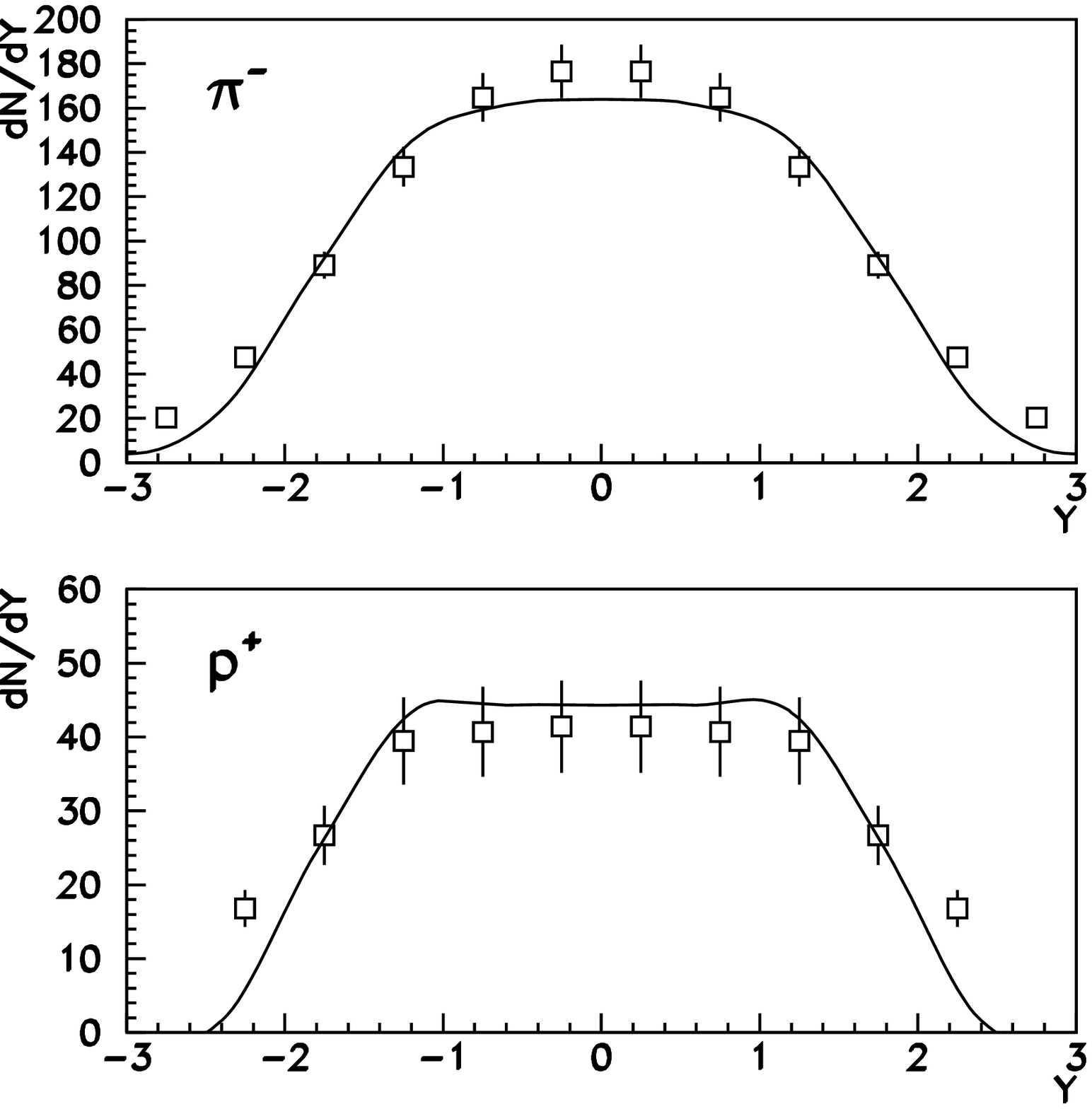}
\vskip 30pt
\begin{minipage}[t]{11.054cm}
{\small {\bf Fig.~1.}
{Rapidity distributions of pions (top) and protons (bottom) from
the extended ALCOR model for
160 AGeV Pb + Pb central collisions.
 The open squares
are the measured and the reflected data points for {\it h-} particle (top)
and {\it '(+) - (-)'} particle (bottom) spectra \cite{PbPb}.
The value of proton multiplicity at mid-rapidity is uncertain.
In more recent publication \cite{PbPbQM96} a small drop was given,
but even the statistical errors are appreciable, yet.} }
\end{minipage}
\end{center}
 
\newpage
 
\begin{center}
\vspace*{14cm}
\includegraphics{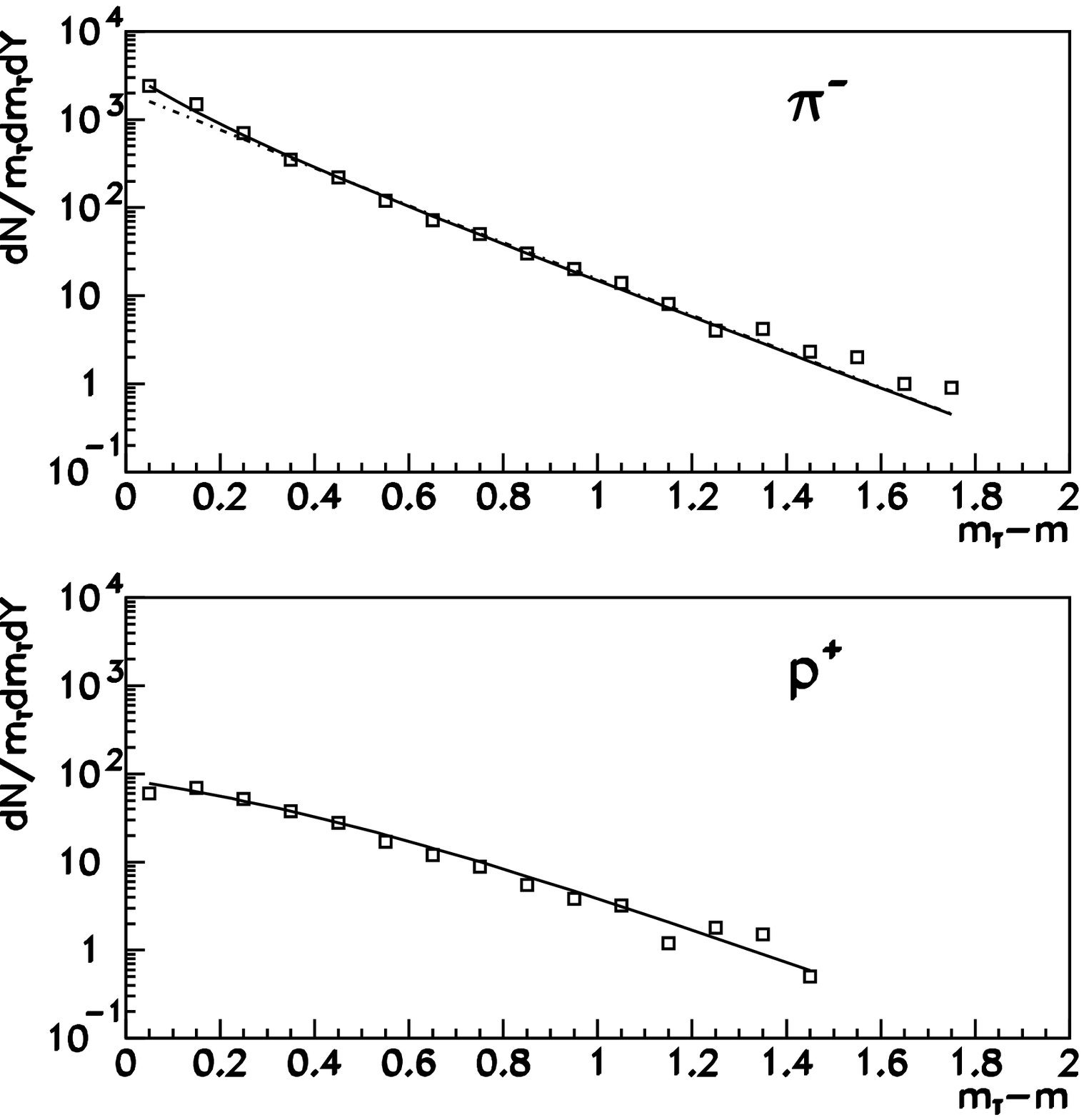}
\vskip 30pt
\begin{minipage}[t]{11.054cm}
{\small {\bf Fig.~2.}
{Transverse mass, $m_T$,
distribution of pions (top) and protons (bottom)
calculated from ALCOR for 160 AGeV Pb+Pb
collisions at mid-rapidity.
The dotted line in the upper figure shows the result obtained
with Boltzmann distribution. The open squares
are the measured values for {\it h-} particle (top)
and {\it '(+) - (-)'} particle (bottom). } }
\end{minipage}
\end{center}
 
\newpage
 
\begin{center}
\vspace*{14cm}
\includegraphics{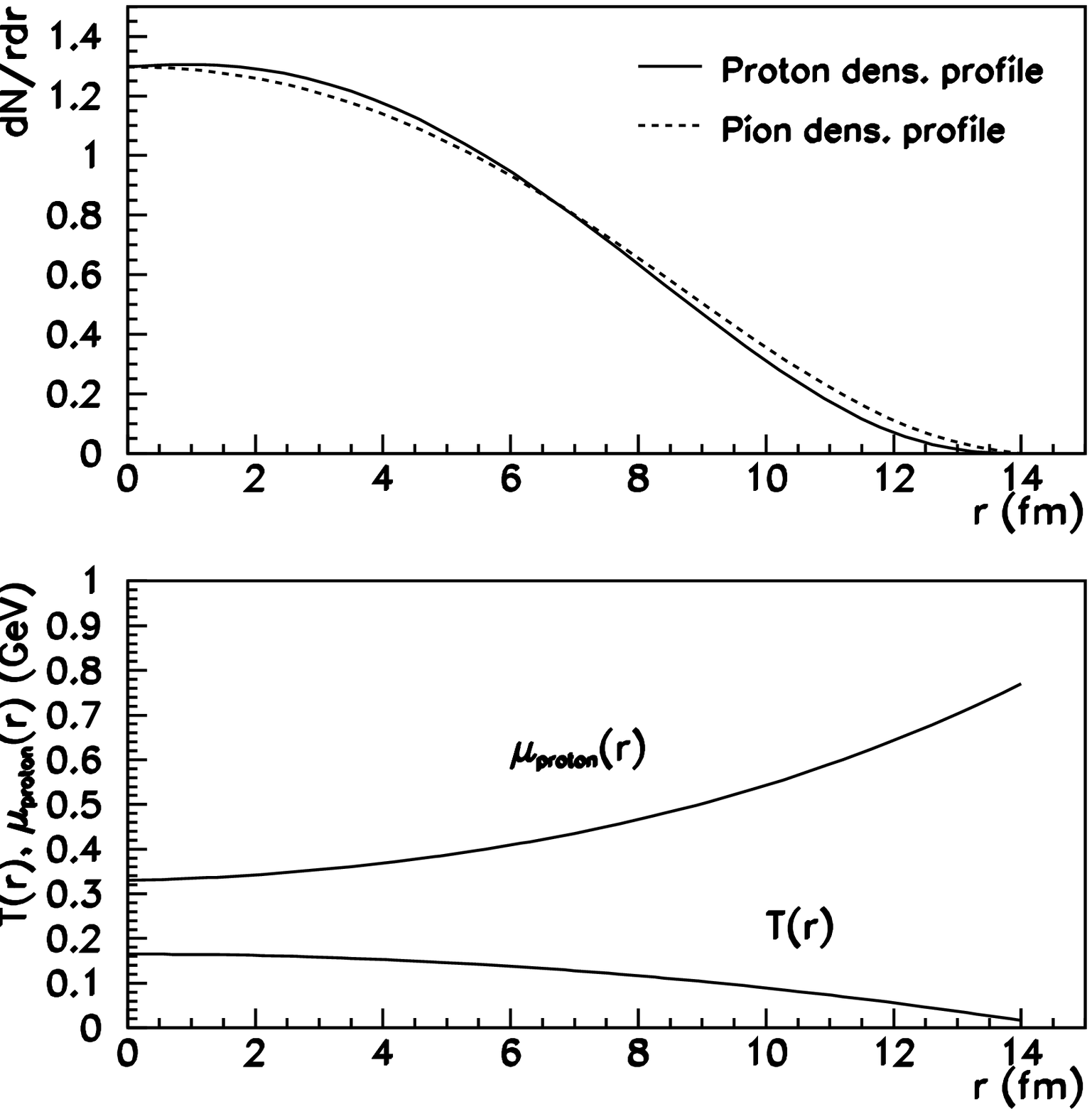}
\vskip 30pt
\begin{minipage}[t]{11.054cm}
{\small {\bf Fig.~3.}
{The approximately similar radial density profiles
for pions and protons (top)
and the common temperature profile, $T(r)$ (bottom)
together with the calculated chemical potential for protons,
$\mu_{proton}$. The pion chemical potential was constant,
$\mu_{\pi}=103 \ MeV$. } }
\end{minipage}
\end{center}
 
\newpage
We would like to point out, that the description of the particle spectrum
from the extended ALCOR model is based on a parameterization of the flow,
density and temperature profiles of the hadronic phase in the time interval
when the hot hadronic matter freezes out.
This parameterization is qualitatively similar to the ones developed in
refs.~\cite{csb95}-\cite{tbs96}, for the description of the single-particle
spectra and the correlation function of the $\pi$ and $K$ mesons,
as has been measured by the NA44 collaboration.
 
Both models confirm that the 
transverse flow profile, the local density profile
and the transverse temperature profile~\cite{csb96}-\cite{tbs96}
may play a crucial role when describing the hadronic spectra.
 
\section*{ Conclusions}

 In this paper we introduced the concept of transchemistry for
the description of hadronization.  The ALCOR model is a good
approximation to the transchemistry description, if the
hadronization is fast.  The ratios of different hadron
multiplicities obtained from ALCOR and from the thermal model --
based on thermal and chemical equilibrium of hadron resonances --
are much closer to each other, than one could expect.  The
rapidity and transverse mass distribution from ALCOR was compared
to the experimental ones from the SPS PB+Pb collisions.  The
agreement is good.
 
Further we have shown that the Bose statistics may play a role in
the low
$m_t$ enhancement in the transverse mass distribution.
 
These results support the claims that in high energy heavy ion
collisions a semi-deconfined state of the matter is formed in Pb
+ PB reactions at CERN SPS energies.

Since it seems  that in the  particle composition there  are only minor,
but systematic  differences between  ALCOR and  thermal models,  refined
analysis tools are expected to play  a critical role in the future  data
analysis: especially $\chi^2$  fits to the  slope parameters and  to the
detailed invariant  momentum distribution  of various  kinds of detected
particles may  be necessary  to clearly  distinguish among  the proposed
models.
 
Note also that the total volume and the inner dynamics characterized  by
the local  temperature, flow  and chemical  potential distributions  are
very  different  in  ALCOR  and  in  the  models assuming global thermal
equilibrium.  The HBT radius parameters  are known to be very  sensitive
to  the  detailed  structure  of  the  flow,  temperature  and  chemical
potential  profiles.   Thus  we  expect that their
evaluation using parameters fitted to the spectrum can possibly  clearly
distinguish between ALCOR  and simple thermal  models.  Further work  is
necessary to evaluate this parameters from the models.

\newpage
 
\section*{ Acknowledgment}
Discussions with M. Gazdzicki, B. M\"uller and R. Stock
are acknowledged.
This work was supported by the National Scientific
Research Fund (Hungary), OTKA
No.T016206 and F4019 as well as
by the U.S. - Hungarian Science and Technology
Joint Fund, No. 378/93.


\end{document}